\documentclass{aa}

\usepackage{graphicx}
\usepackage{txfonts}
\usepackage{amsmath}
\usepackage{natbib}

\usepackage{color}
\usepackage{multirow}
\usepackage{mathtools}
\usepackage{epstopdf}

\usepackage{hyperref}
\usepackage{url}
\usepackage[dvipsnames]{xcolor}

\usepackage{gensymb}
\usepackage{appendix}
\usepackage{verbatim}
\usepackage{enumitem}
\usepackage{stfloats}

\hypersetup{colorlinks=true, citecolor=MidnightBlue, linkcolor=MidnightBlue}

\begin{document}

\title{The imprint of synchronized-pair seeding of LRDs with direct collapse black holes}
\titlerunning{Synchronized-pair seeding in LRDs}

\author{F. Ziparo\inst{1}, S. Carniani\inst{1}, E. Parlanti\inst{1}, B. Trefoloni\inst{1}, S. Zamora\inst{1}, A. Ferrara\inst{1}, S. Gallerani\inst{1}}
\authorrunning{F. Ziparo et al.}
\institute{Scuola Normale Superiore, Via Consoli del Mare, Pisa, Italy
           \\ \email{francesco.ziparo@sns.it}}
\date{Received ---; accepted ---}

\abstract
{The frequent detection of UV-bright companions around little red dots (LRDs) has motivated a possible connection between their environments and black hole (BH) seed formation. We investigate this connection using 25 LRD systems with virial BH mass estimates and UV-bright companions, including four spectroscopically confirmed pairs. We estimate the incident companion UV specific intensity ($J_{\rm UV}$) from the companion luminosities and projected separations. The data suggest a positive mass--intensity relation, $\log_{10}(M_{\rm BH}/M_\odot)=0.88\log_{10}[J_{\rm UV}/ ({\rm erg\,s^{-1}\,cm^{-2}\,Hz^{-1}\,sr^{-1}})]+22.34$.
We explore an interpretation within the synchronized-pair scenario, in which Lyman--Werner (LW) radiation from a neighboring galaxy enables direct-collapse BH formation. In our timing prescription, stronger irradiation allows earlier seed formation and leaves more time for subsequent BH growth. Assuming an exponentially rising companion star-formation history and exponential BH growth, the model predicts a power-law mass--intensity relation whose slope measures the ratio of the companion luminosity and BH mass e-folding times, $\tau_\star/\tau_{\rm BH}$. Its normalization provides conditional constraints on the seed mass. For a critical LW intensity $J_{\rm crit}=10^3\,J_{21}$, we infer $\log_{10}(M_{\rm seed}/M_\odot)=6.72^{+0.29}_{-0.39}$, above typical direct-collapse seed masses. Adopting $J_{\rm crit}=50\,J_{21}$, or assuming that the LRD BH masses are on average overestimated by $\sim1.5$ dex, lowers the inferred logarithmic seed mass to approximately $5.6$ or $5.2$, respectively, easing this tension. The tentative trend and proposed timing model provide a framework for testing whether LRDs trace an early phase of direct-collapse BH evolution. Larger samples and explicit tests of luminosity-dependent selection effects are needed to establish whether the relation retains information about BH seed formation.}

\keywords{Supermassive black holes -- galaxies: high-redshift -- galaxies: nuclei}

\maketitle

% ============================================================
\section{Introduction}
\label{sec:introduction}

The James Webb Space Telescope (\textit{JWST}) has revealed a population of broad-line active galactic nuclei (AGN) at $z\gtrsim4$, extending the study of early black hole (BH) growth to luminosities substantially below those of the quasars previously accessible
at these redshifts. Among this broad line AGN population, "Little Red Dots" (LRDs) arguably represent the most intriguing subsample: these sources are characterized by compact rest-frame optical emission and red optical continua, often accompanied by a blue rest-frame ultraviolet (UV) component \citep{Kocevski2023,Greene2024,Matthee2024}.  The detection of broad Balmer lines in many LRDs has motivated interpretations in which a central accreting BH might contribute to their emission. Their spectral properties nevertheless differ from those of conventional AGN, indicating that the surrounding gas can play an important role in shaping the observed emission \citep{Taylor2025, deGraaff2026}.

Absorption features in broad Balmer lines and pronounced Balmer breaks observed in some LRDs have motivated models in which an accreting BH is embedded in dense gas \citep{InayoshiMaiolino2025, Kokorev2026, Naidu2026, Gentile2026}.  Absorption, scattering and reprocessing in this gas can modify both the continuum and emission-line profiles. In particular, \citet{Rusakov2026, Modelevsky2026} have explored models in which BHs are embedded in dense, optically thick gas envelopes, with electron scattering contributing substantially to the line broadening and radiative reprocessing in the surrounding gas shaping the optical continuum. Such gas-enshrouded systems have been proposed to represent an early phase of BH growth \citep{Inayoshi2025, Begelman2026, Pacucci_Ferrara2026}. Connecting LRDs to a specific seeding mechanism, however, requires establishing whether current LRD properties retain information about the conditions under which their BHs initially formed.

The large BH-to-stellar mass ratios inferred for several LRDs provide a  motivation for considering a massive-seed origin \citep{Maiolino2024, Juodzbalis2026, Pacucci_Ferrara2026, Whalen2026}.
Under the conventional virial interpretation of their broad Balmer lines, these systems appear to host BHs that are overmassive relative to the local BH--host stellar mass relation  \citep{Greene2024, Jones2026}, although observational selection
effects may contribute to the inferred offset \citep{Ziparo2026}.
Although elevated mass ratios do not uniquely identify the seeding mechanism, a massive seed formed before substantial stellar assembly could naturally produce such a configuration, with a BH mass initially comparable to, or exceeding, the host's stellar mass.
Models of massive-seed formation show that elevated BH-to-stellar mass ratios can persist during subsequent evolution, potentially retaining information about the initial conditions \citep{ScogginsHaiman2024}.
These considerations motivate models linking LRDs to the formation and early growth of massive BH seeds in systems with little prior star formation \citep{Pacucci_Ferrara2026,Whalen2026}.

Direct collapse can produce massive seeds with initial masses of approximately $10^4$--$10^5\,M_\odot$ \citep[DCBHs;][]{BrommLoeb2003,Inayoshi2020}. A possible realization of this mechanism is the synchronized-pair scenario, in which two neighbouring,
initially pristine haloes reach the atomic-cooling threshold within a short time interval \citep{Visbal2014}. Star formation in the first halo produces a Lyman--Werner (LW) radiation field that dissociates molecular hydrogen in the second. Suppression of molecular
cooling can prevent the gas from cooling below the atomic-cooling regime, limiting fragmentation and maintaining the inflow required to form a massive central object. However, the irradiating stellar population also produces ionizing photons and metals, which can disrupt these conditions. Successful collapse therefore requires a suitable combination of separation, irradiation history and synchronization, allowing the target to collapse before photoevaporation or metal enrichment becomes effective \citep{Regan2017, Yue2017}.

The identification of UV-bright components near LRDs has provided an observational basis for investigating this scenario. \citet{Baggen2026} compiled 83 LRDs with \textit{JWST} imaging and identified candidate companions in approximately $43\%$ of
the systems, at projected separations ranging from sub-kpc to several-kpc scales. Assuming a stellar origin for the companion emission, their component-resolved photometry implies that many LRDs are exposed to LW intensities comparable to those required in
direct-collapse models. The observed configurations are therefore consistent with the presence of potential external irradiating sources. Physical association between the components remains yet to be established spectroscopically for much of the sample.

\begin{figure*}[t]
    \centering
    \includegraphics[width=0.8\textwidth]{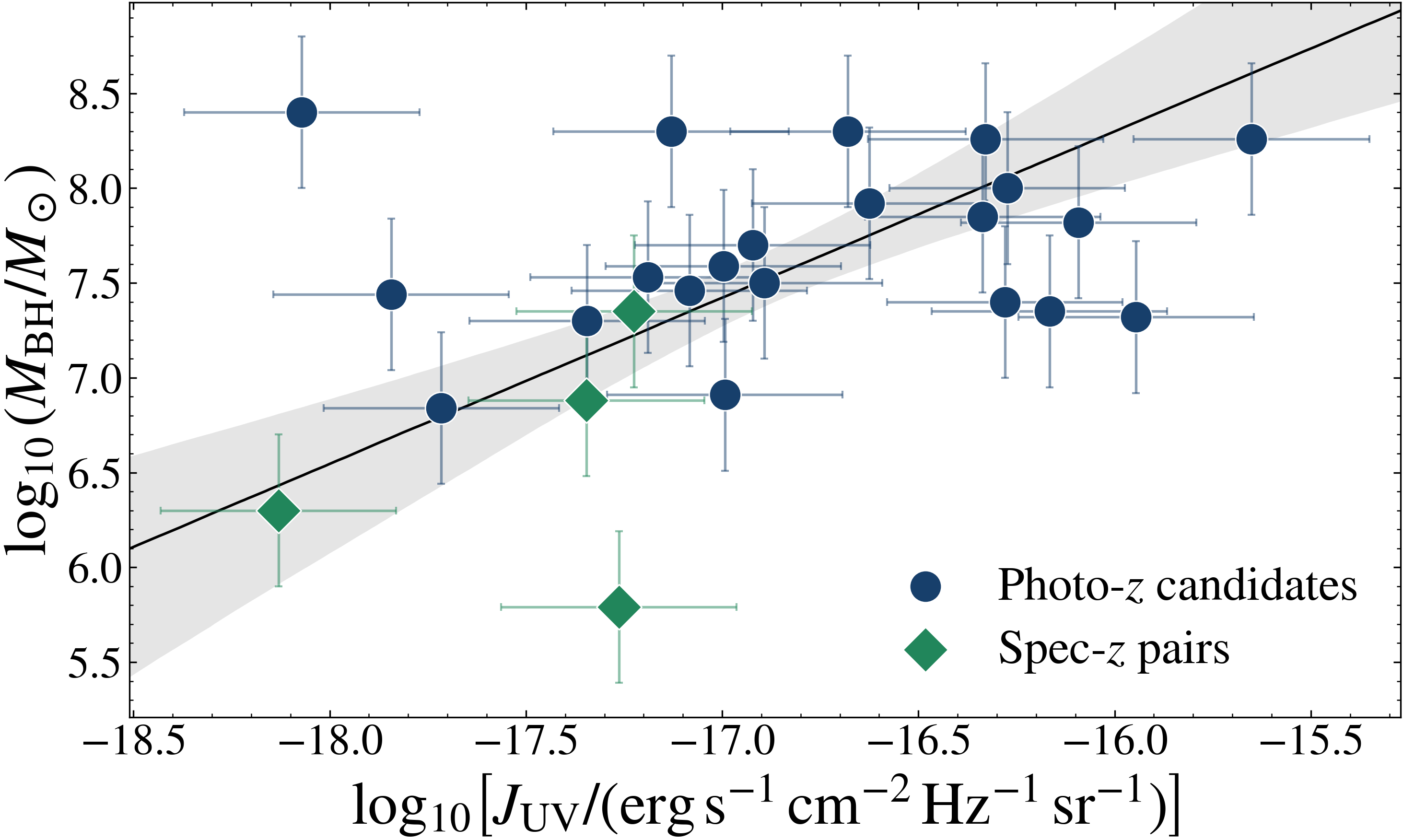}
    \caption{BH mass as a function of the estimated incident
    companion UV specific intensity. Green diamonds identify
    spectroscopically confirmed companion associations;
    blue circles denote photometric candidate associations.
    Error bars show the adopted uncertainties of $0.4$~dex
in BH mass and $0.3$~dex in UV intensity, with effective
separations held fixed.
The black line and grey band show the pointwise median
and 16th--84th percentiles of the relations sampled from
the joint posterior of $\alpha$ and $\beta$.}
    \label{fig:relation}
\end{figure*}

Here we investigate whether the BH masses of LRDs correlate with the incident UV intensities inferred from their companions, and explore a possible interpretation within the synchronized-pair scenario. The paper is organized as follows. Section~\ref{sec:results} describes the sample and observables and presents the observed mass--intensity relation. Section~\ref{sec:interpretation} develops its analytical interpretation within the synchronized-pair scenario and examines the implications for seed masses and BH growth. Section~\ref{sec:discussion} discusses the model assumptions, sources of scatter and implications for LRDs without detected companions. The statistical significance and robustness tests are presented in Appendix~\ref{app:robustness}.

% ============================================================
\section{Observed mass--intensity relation}
\label{sec:results}

Our sample comprises 25 spectroscopically confirmed LRDs at $2.27<z<6.99$ with UV-bright companions at projected separations of up to a few kiloparsecs. The companions have photometric redshifts consistent with those of the corresponding LRDs. Four pairs are also spectroscopically confirmed, with relative line-of-sight velocity offsets  $\lesssim 100~{\rm km\,s^{-1}}$. Twenty-four systems are drawn from the companion catalogue of \citet{Baggen2026}, with one additional system from Zamora et al.\ (\textit{in preparation}). All LRDs in our sample show broad Balmer lines associated with broad-line region (BLR) emission. The BH masses used in this work are based on single-epoch virial estimates reported by \citet{Greene2024}, \citet{Juodzbalis2024}, \citet{Matthee2024}, \citet{Taylor2025} and \citet{Baccus2026}. Table~\ref{tab:sample}  lists the LRDs used in this work and reports their main properties.

Figure~\ref{fig:relation} shows the BH masses of the selected LRDs as a function of the estimated mean UV specific intensity ($J_{\rm UV}$) produced by their companions at the locations of the LRDs. Green diamonds and blue circles identify spectroscopically confirmed and candidate associations, respectively.

We convert the companion UV absolute magnitudes reported in the catalogue of \citet{Baggen2026} into UV specific luminosities,
$L_{\rm UV,obs}$, using the AB magnitude definition. We then estimate the UV specific intensity at the LRD location as $J_{\rm UV}=L_{\rm UV,obs}/(16\pi^2d^2)$, approximating the true physical separation $d$ by the projected separation. For systems with multiple companions, we use the total companion UV specific luminosity and the effective projected separation defined by \citet{Baggen2026}. This is the separation at which a single source with the total companion luminosity would reproduce the combined radiation field from all companions. We adopt a fiducial uncertainty of 0.3 dex in $J_{\rm UV}$ for all systems to account for measurement errors and the uncertainty introduced by approximating the true physical separations by their projected values. For the BH mass estimates we adopt  an uncertainty of $0.4$~dex for each BH mass to account for uncertainties in the calibration of the virial mass estimators \citep{Greene2005}. 

Figure \ref{fig:relation} shows a positive correlation between BH mass and the incident companion UV intensity. We quantify the mass--intensity trend using Spearman's rank correlation, obtaining $\rho_{\rm S}=0.38$ with $p_{\rm perm}=6.12\times10^{-2}$. The positive coefficient indicates that systems exposed to stronger companion UV irradiation generally host more massive BHs. This provides evidence for a positive correlation between BH mass and companion UV intensity.

We thus fit the data with a linear relation in logarithmic space:
\begin{equation}
\log_{10}\!\left(\frac{M_{\rm BH}}{M_\odot}\right)=\beta+\alpha\log_{10}\!\left[\frac{J_{\rm UV}}{{\rm erg\,s^{-1}\,cm^{-2}\,Hz^{-1}\,sr^{-1}}}\right],
\label{eq:observed_flux_relation}
\end{equation}
yielding the best-fit results $\alpha=0.88^{+0.40}_{-0.28}$ and $\beta=22.34^{+6.80}_{-4.83}$, where the quoted intervals enclose the 16th--84th percentiles of the marginal distributions. The slope is consistent with a linear scaling; its central value corresponds to an increase in BH mass by a factor of approximately eight per decade in companion UV intensity.

To assess the statistical significance of the observed mass--intensity trend, we also compare a free-slope model with a flat relation ($\alpha=0$), allowing for intrinsic scatter in both models. A parametric bootstrap conditional on the observed UV intensities
gives $p_{\rm boot}=0.048$, providing modest evidence against a flat relation. We assess the stability of the trend under substantial changes in sample composition by repeating the orthogonal Bayesian fit for $10^4$ random subsamples, each obtained by removing roughly
one-third of the systems (7 out of 25, $28\%$). The posterior median slope is positive in 9,989 trials. Across subsamples, the median
of these slopes is $0.92$, with 16th--84th percentiles of $[0.59,1.14]$. The positive slope is therefore stable under these changes in sample composition. The statistical procedures and their assumptions are described in Appendix~\ref{app:robustness}.

% ============================================================

\section{Physical interpretation}
\label{sec:interpretation}

The relation measured in Sec.~\ref{sec:results} links BH mass with companion irradiation at the epoch of observation. We consider the synchronized-pair scenario as a possible interpretation and examine which seeding conditions and growth histories could reproduce the inferred slope and normalization.

In this scenario, LW radiation from a nearby star-forming galaxy suppresses molecular hydrogen cooling in the target halo, allowing its gas to remain on the atomic-cooling track and creating conditions favourable for direct collapse \citep{Shang2010,Visbal2014,Regan2017}. Beyond enabling this process, the irradiation history may also influence when the seed forms and hence the time available for subsequent BH growth. For a companion with a rising luminosity, the critical LW intensity can be reached earlier at smaller separations. Provided that the target already satisfies the other conditions for direct collapse and remains irradiated throughout collapse, this can translate into earlier seed formation. For otherwise comparable seed masses and accretion histories, the additional time available for growth can therefore produce a larger BH mass at observation.

\begin{figure}[h!]
    \centering
    \includegraphics[width=1\linewidth]{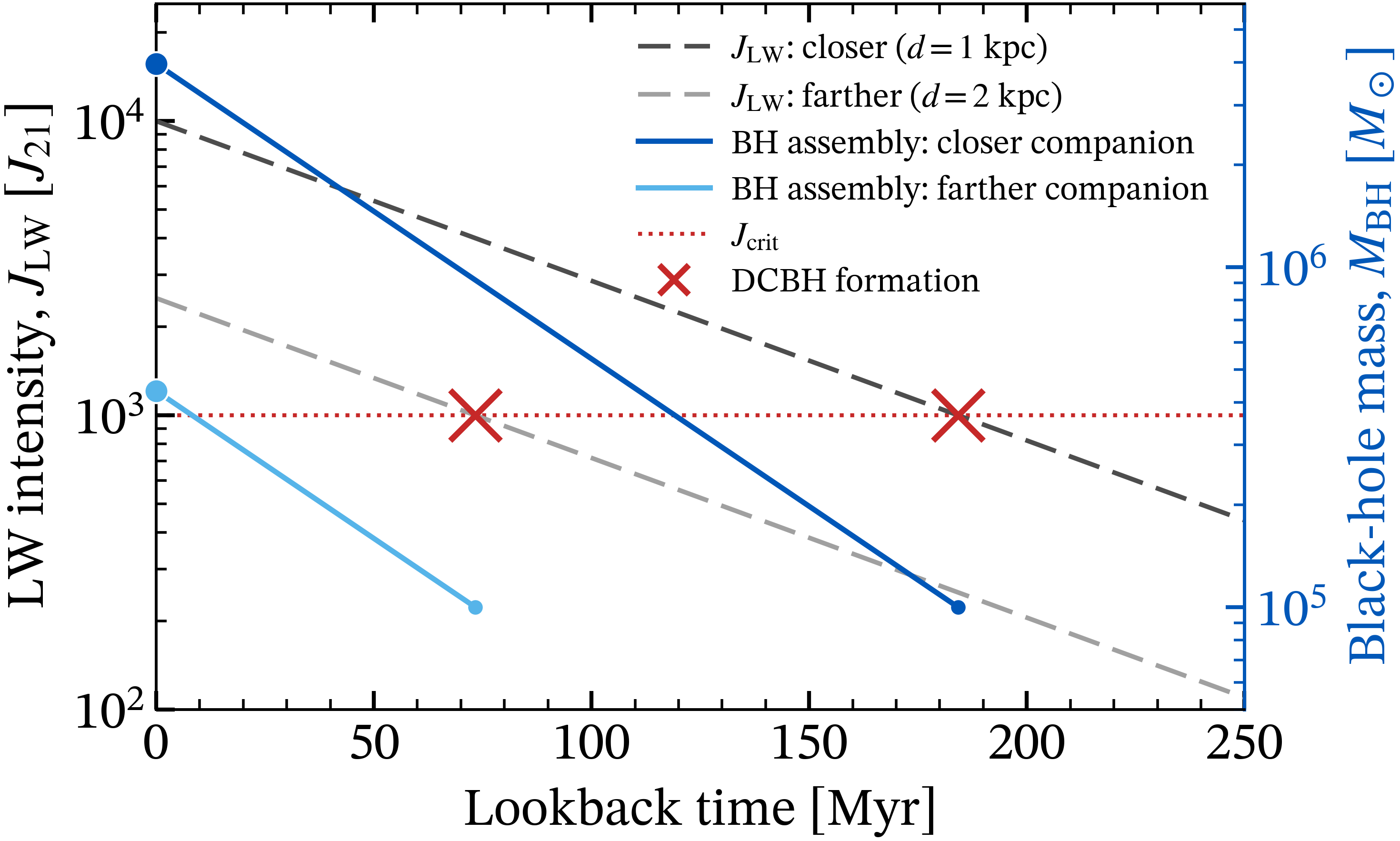}
\caption{Schematic illustration of the proposed connection between
companion separation, seed formation time, and BH mass. Dashed curves
show the LW intensity for the same companion luminosity history at
separations of $1$ and $2\,\mathrm{kpc}$ (dark and light grey,
respectively; left axis). The dotted red line marks the illustrative
threshold $J_{\rm crit}=10^{3}\,J_{21}$, and red crosses indicate the
assumed seed formation epochs. Solid blue curves show subsequent
growth from $M_{\rm seed}=10^{5}\,M_\odot$ at the same exponential continuous
growth rate (right axis). The closer system forms its seed earlier
and reaches a larger BH mass at $z=5$. Cosmic time increases from
right to left.}
\label{fig:schematic}
\end{figure}

Figure~\ref{fig:schematic} illustrates this timing argument for two systems with the same increasing companion luminosity history at fixed physical separations of $d=1$ and $2$~kpc. The closer target reaches the critical intensity earlier and therefore forms its seed earlier, leaving more time for subsequent BH growth. The parameters adopted in this schematic picture are illustrative.

We now formulate this timing argument analytically and examine how it can account for the measured mass--intensity relation. We adopt a simple timing prescription in which seed formation occurs when the LW intensity $J_{\rm LW}$ reaches a critical threshold $J_{\rm crit}$. We express LW intensities in units of $J_{21}\equiv10^{-21}\, {\rm erg\,s^{-1}\,cm^{-2}\,Hz^{-1}\,sr^{-1}}$. We define $t$ as the lookback time relative to the epoch of observation and, adopting an irradiation history $J_{\rm LW}(t)$, the seed-formation lookback time $\bar t$ satisfies $J_{\rm LW}(\bar t)=J_{\rm crit}$. Assuming isotropic emission and negligible attenuation, the LW specific intensity at the expected BH location due to radiation from the companion is
\begin{equation} 
J_{\rm LW}(t) = \frac{\kappa L_{\rm UV}(t)}{16\pi^2 d^2}
\label{eq:incident_uv_flux}
\end{equation}
where $d$ is the physical separation, held fixed in this model, $L_{\rm UV}(t)$ is the companion's UV specific luminosity as a function of lookback time, and $\kappa=L_{\rm LW}/L_{\rm UV}$ is the adopted LW-to-UV specific luminosity ratio.

The UV luminosity of a young galaxy depends on its recent star-formation history. As a first-order approximation, we assume that the UV luminosity follows
the star-formation rate and adopt an exponentially rising history,
which provides an acceptable fit to most of the reconstructed
high-redshift SFHs examined in Appendix~\ref{app:sfh_tau}. In terms of lookback time $t$, this gives $L_{\rm UV}(t)\propto\mathrm{SFR}(t)\propto e^{-t/\tau_\star}$, where $\tau_\star$ is the characteristic star formation e-folding time. Therefore, we can write that  
\begin{equation} 
L_{\rm UV}(t)= L_{\rm UV,obs}e^{-t/\tau_\star}
\label{eq:uv_luminosity}
\end{equation}
where $L_{\rm UV,obs}$ is the companion luminosity at the time of observation. Applying the threshold condition $J_{\rm LW}(\bar t)=J_{\rm crit}$ to Eqs.~\ref{eq:incident_uv_flux} and~\ref{eq:uv_luminosity} gives
 \begin{equation} 
\bar t= \ln\left(\frac{L_{\rm UV,obs}}{16\pi^2 d^2}\frac{\kappa}{J_{\rm crit}}\right)\tau_\star
\label{eq:seeding_time}
\end{equation}

Assuming that the BH grows through gas accretion at an average constant Eddington ratio ($\lambda_{\rm Edd}$) and fixed radiative efficiency ($\varepsilon$) from seed formation to observation, its mass at the observed epoch is:
\begin{equation}
M_{\rm BH,obs}=M_{\rm BH,seed}e^{\bar t/\tau_{\rm BH}},
\label{eq:bh_growth}
\end{equation}
where $\tau_{\rm BH}$ is the BH mass e-folding time. 
Combining equations \ref{eq:seeding_time} and \ref{eq:bh_growth}, we find: 
\begin{equation}
\log_{10}\left(\frac{M_{\rm BH,obs}}{\rm M_\odot}\right)=\frac{\tau_\star}{\tau_{\rm BH}}\log_{10}\left(\frac{L_{\rm UV,obs}}{16\pi^2 d^2}\frac{\kappa}{J_{\rm crit}}\right)+\log_{10}\left(\frac{M_{\rm BH,seed}}{\rm M_\odot}\right)
\label{eq:timing_mass_intensity}
\end{equation}

Under these assumptions, the model yields a linear relation in logarithmic space between the observed BH mass of an LRD ($M_{\rm BH,obs}$) and the mean UV specific intensity produced by its companion at the BH location ($L_{\rm UV,obs}/(16\pi^2d^2)$). The slope, $\alpha_{\rm model}=\tau_\star/\tau_{\rm BH}$, reflects the ratio of the characteristic timescales for companion luminosity evolution and BH mass growth. The intercept depends jointly on the initial BH seed mass and the critical LW intensity, together with the adopted LW-to-UV luminosity ratio: 
\begin{equation}
\beta_{\rm model} = \frac{\tau_\star}{\tau_{\rm BH}}\log_{10}\frac{\kappa}{J_{\rm crit}}+\log_{10}(M_{\rm BH,seed}/M_\odot).
\label{eq:intercept}
\end{equation}

\begin{figure}
    \centering
    \includegraphics[width=1\linewidth]{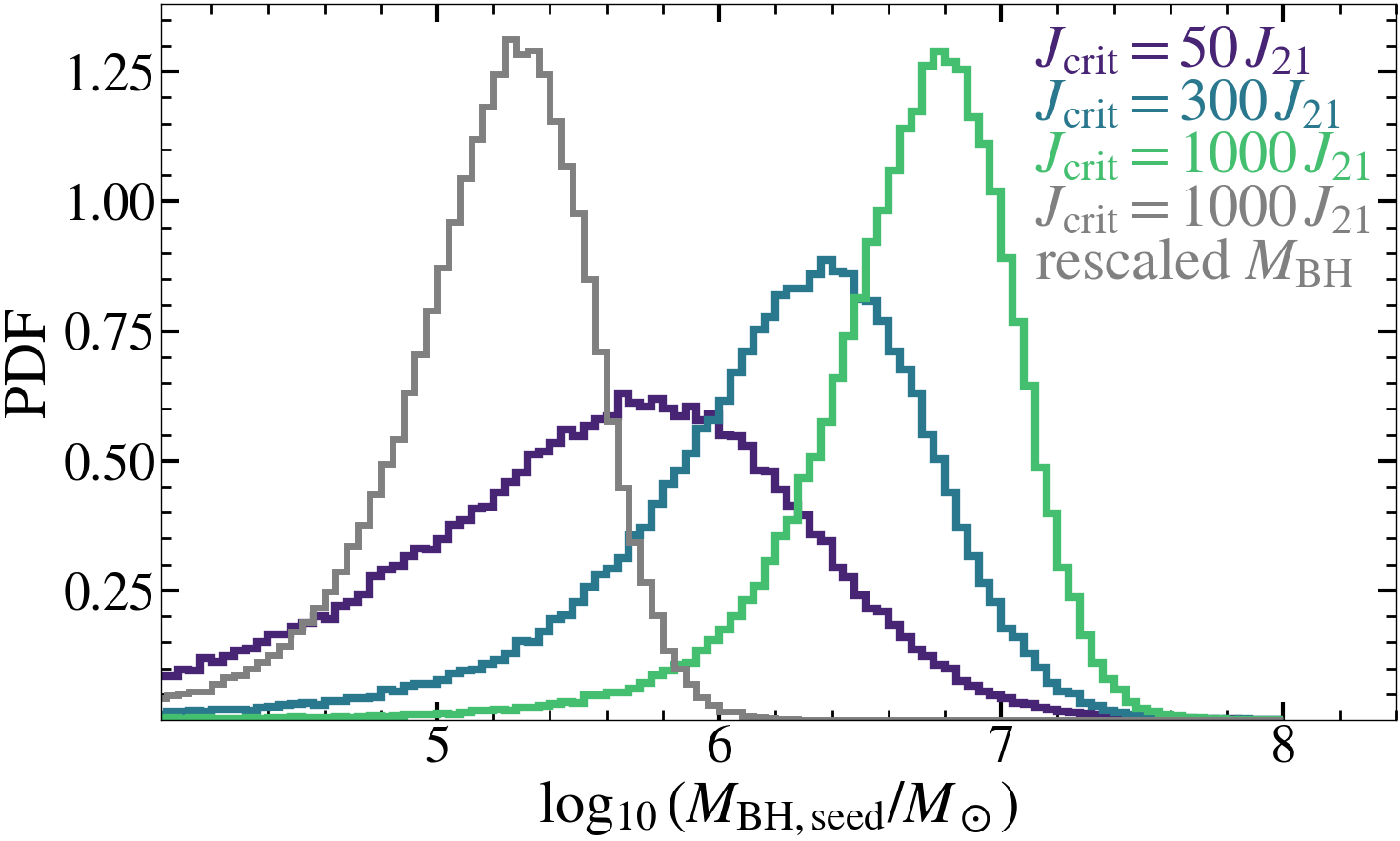}
    \caption{Posterior probability densities of the logarithmic BH seed mass inferred from the joint posterior of the slope and intercept of the mass--UV intensity relation under the analytical model. The coloured curves correspond to critical LW intensities of $J_{\rm crit}=50$, 300 and $1000\,J_{21}$ (purple, blue and green, respectively), adopting a fixed LW-to-UV specific luminosity ratio $\kappa=0.63$. The grey curve shows the result for $J_{\rm crit}=1000\,J_{21}$ after uniformly reducing the inferred BH masses by $1.5$~dex, illustrating the effect of a possible systematic overestimate of the observed masses.}
    \label{fig:seed_mass}
\end{figure}

The normalization constrains the seed mass for an adopted critical LW intensity and LW-to-UV conversion factor, $\kappa$. We calibrate $\kappa$ using the companion UV magnitudes and estimated LW intensities tabulated by \citet{Baggen2026}. We approximate this conversion by $M_{\rm LW}\simeq M_{\rm UV}+0.5$. For AB magnitudes, this implies $\kappa=L_{\nu,\rm LW}/L_{\nu,\rm UV}=10^{-0.4(M_{\rm LW}-M_{\rm UV})}\simeq0.63$. Figure~\ref{fig:seed_mass} shows the posterior distributions of the BH seed mass obtained for different assumed critical LW intensities required for direct collapse. Adopting the commonly used threshold $J_{\rm crit}=1000\,J_{21}$, we infer $\log_{10}(M_{\rm BH,seed}/M_\odot)=6.72^{+0.29}_{-0.39}$, above the seed masses predicted by direct-collapse models \citep{Ferrara2014}. Lowering the threshold to $J_{\rm crit}=300\,J_{21}$ and $J_{\rm crit}=50\,J_{21}$ yields $\log_{10}(M_{\rm BH,seed}/M_\odot)=6.26^{+0.43}_{-0.59}$ and $\log_{10}(M_{\rm BH,seed}/M_\odot)=5.58^{+0.64}_{-0.90}$, respectively. These estimates are closer to theoretical expectations, although
their central values exceed $10^5\,M_\odot$ by approximately
1.3 and 0.6 dex, respectively. We discuss this tension with direct-collapse models in greater detail in the following section.

The fitted slope instead constrains the ratio between the companion luminosity and BH mass e-folding times, $\alpha=\tau_\star/\tau_{\rm BH}=0.88^{+0.40}_{-0.28}$, indicating that these timescales are comparable. Within this framework, reconstructing the star-formation histories of the companions could provide an independent constraint on
$\tau_\star$. Indeed, combining this constraint with the measured slope would allow us to infer the effective BH growth timescale, $\tau_{\rm BH}$, without requiring independent estimates of the Eddington ratios or radiative efficiencies of the LRDs.

We note that two of the spectroscopically confirmed companions have been recently analyzed by \citet{PacucciRinaldi2026} and  \citet{Ji2026} and found that they are characterized by a rising star-formation history with a stellar age $<50~{\rm Myr}$ that yield an upper limit on $\tau_{\rm BH}<60~{\rm Myr}$ in our model. In particular, if the rising star-formation histories of the companions have an e-folding time $\tau_\star\simeq20$~Myr during the rising phase of the last burst, we obtain $\tau_{\rm BH}\simeq23$~Myr. For continuous accretion, the corresponding Eddington ratio is
$\lambda_{\rm Edd}=t_{\rm E}\varepsilon/
[(1-\varepsilon)\tau_{\rm BH}]$, where $t_{\rm E}\simeq450$~Myr. Adopting $\varepsilon=0.05$, the inferred growth timescale corresponds to $\lambda_{\rm Edd}\simeq1$. Such an e-folding time thus points to a regime of rapid BH growth fed by extremely dense gas. The associated high mass inflow rates can bring the accretion flow into the super-critical regime, where photon trapping reduces the radiative efficiency and limits the emergent luminosity to values of order $L_{\rm Edd}$ \citep{Madau2014}.  The same gas reservoir could also obscure the central source during the early growth of the DCBH. In the reference calculations discussed by \citet{Pacucci_Ferrara2026}, accretion progressively consumes this reservoir until a final radiation-driven event expels the remaining gas after approximately $100$--$200$~Myr. For the seed masses implied by different values of \(J_{\rm crit}\), the elapsed growth times can then be compared with the duration of the obscured phase. For the central fitted slope and normalization, $J_{\rm crit}\simeq170\,J_{21}$ corresponds to $M_{\rm BH,seed}\simeq10^6\,M_\odot$. With $\tau_{\rm BH}\simeq23$~Myr, this seed would reach $M_{\rm BH,obs}=10^8\,M_\odot$ approximately 106~Myr after formation. A lower threshold, $J_{\rm crit}\simeq12\,J_{21}$, corresponds to $M_{\rm BH,seed}\simeq10^5\,M_\odot$ and an elapsed time of approximately 159~Myr, although this threshold lies below the range adopted above. We note that the short inferred ages leave less time for the pair separation to evolve, lending support to our fixed-separation approximation.

% ============================================================
\section{Discussion}
\label{sec:discussion}

The observed mass--intensity relation provides an environmental constraint on the formation and growth of BHs in LRDs. Its interpretation within the synchronized-pair scenario requires connecting the measured BH and companion properties to the conditions at seed formation and their subsequent evolution. We discuss the physical and observational factors that affect this interpretation and its implications for the wider LRD population.

\subsection{Halo synchronization and the LW threshold criterion}
Identifying LW threshold crossing with seed formation assumes synchronization between the growth of the target halo and the rise of the irradiating field. If the target reaches the atomic-cooling threshold before molecular cooling is suppressed, gas fragmentation and star formation can occur. Subsequent metal enrichment can promote further cooling and fragmentation, preventing the pristine collapse considered here \citep{Visbal2014,Regan2017}. Joint modeling of halo assembly and irradiation is therefore required to establish whether and when this seeding channel remains viable in individual systems \citep{Chon2016}.

\subsection{Orbital evolution and the fixed-separation approximation}
In our analytical derivation, we assume that the physical separation between the target halo and its irradiating companion remains fixed over time. Orbital motion and interactions can change this separation, so its value during seed formation may differ from that at the observed epoch.  Even at observation, the physical separation is only partly constrained by imaging, since $d_{\rm eff}$ is inferred from the configuration projected on the sky and therefore depends on the viewing angle. Orbital evolution and projection can thus contribute to the scatter in the observed mass--intensity relation. If the LRD and its companion form a gravitationally bound system, their separation can change both along the orbit and through orbital decay. For low-eccentricity orbits and elapsed times shorter than the orbital decay timescale, the separation remains approximately constant, supporting our fixed-separation approximation. However, if orbital decay is appreciable, the larger separations in the past would shift seed formation time closer to observation, leaving less time for BH growth.

\subsection{Critical LW intensity and virial BH mass systematics}
The inferred seed masses depend on both the adopted critical LW intensity and the BH mass estimates. A systematic downward revision of the BH masses or lower values of $J_{\rm crit}$ would both reduce the inferred seed masses, easing the tension with predictions from DCBHs formation models. The grey curve in Fig.~\ref{fig:seed_mass} illustrates a uniform downward revision of the BH masses by $1.5$~dex, motivated by the lower mass estimates reported by \citet{Rusakov2026}. For our fiducial assumption of $J_{\rm crit}=1000\,J_{21}$, this reduces the inferred seed mass to $\log_{10}(M_{\rm seed}/M_\odot)\simeq5.2$, closer to the theoretical
predictions for the formation of DCBHs ($M_{\rm seed}\sim10^5\,M_\odot$). Similarly, adopting $J_{\rm crit}\simeq20\,J_{21}$ with the original BH mass estimates would yield comparable seed masses, although this threshold lies below the range considered above.

\subsection{Origin of the offset UV emission}
The interpretation of the mass--intensity relation also depends on
the origin of the offset UV emission. Although our sample includes spectroscopically confirmed companion galaxies, some UV components could arise from nebular emission from star-free gas clouds photoionized by the nearby LRD \citep{chen2025, Ji2026}. Within our sample, we find no clear correlation between the companion UV luminosity and the estimated incident radiation field from the LRD, traced by $L_{\rm H\alpha,LRD}/(4\pi d_{\rm eff}^{2})$. Here, the LRD H$\alpha$ luminosity serves as a proxy for the AGN bolometric luminosity, due to the uncertain applicability of standard AGN bolometric corrections to LRDs. Furthermore, explaining the observed companion UV luminosities ($\langle M_{\rm UV}\rangle=-19.1$) with nebular emission alone requires high ionizing escape fractions from the LRDs in many systems (see Appendix~\ref{app:photoionization}). Taken together, these results disfavor external photoionization as the dominant origin of the offset UV emission across our sample. Although this interpretation remains viable for individual systems, deeper spectroscopic observations will be key to assessing its overall validity.

\subsection{Implications for the LRD population}
The presence of LRDs without detected UV-bright companions raises the question of whether synchronized seeding contributes to the entire population or to a subset. In the compilation of \citet{Baggen2026}, approximately $43\%$ of the 83 LRDs have an identified companion; including tentative associations increases this fraction to about $72\%$. The absence of a detected companion does not necessarily imply
the absence of an irradiating source during seed formation. A companion may remain below the detection limit or unresolved from the red component. Subsequent evolution may also remove a distinct UV-bright component through a merger or a decline in its UV luminosity \citep{Baggen2026,PacucciRinaldi2026}. The observed companion fraction therefore depends on both observational completeness and the duration for which a descendant retains a detectable companion. Different seed origins may also contribute to the LRD population. The compact, red appearance could correspond to a phase of accretion and obscuration shared by BHs originating from distinct channels, including stellar remnants \citep{MadauRees2001} and
runaway stellar collisions \citep{DevecchiVolonteri2009}. Companion non-detection alone cannot distinguish these alternative origins from the evolved descendants of synchronized pairs. Our analysis tests the mass--intensity relation in systems with identified companions and available BH mass estimates. It does not determine the contribution of synchronized seeding to the full LRD population. Constraining this contribution requires companion completeness estimates and comparisons of BH, stellar and gas properties between systems with and without detected companions.

\section{Summary and conclusions}

We have examined the relation between BH mass and the incident UV intensity inferred from nearby UV-bright companions in 25 LRD systems, including four spectroscopically confirmed pairs. Our main results are:

\begin{itemize}

\item Systems exposed to stronger companion UV intensities tend to host more massive BHs. We find a linear relation in logarithmic space, $\log_{10}(M_{\rm BH}/M_\odot)=\beta+\alpha\log_{10}[J_{\rm UV}/({\rm erg\,s^{-1}\,cm^{-2}\,Hz^{-1}\,sr^{-1}})]$ with $\alpha=0.88^{+0.40}_{-0.28}$, and $\beta=22.34^{+6.80}_{-4.83}$.

\item Within our analytical synchronized-pair model, stronger companion irradiation allows the critical LW intensity to be reached earlier, leaving more time for subsequent BH growth. The slope of the mass--intensity relation constrains the ratio of the companion UV
luminosity and BH mass e-folding times, $\alpha=\tau_\star/\tau_{\rm BH}$. An independent estimate of either timescale would therefore constrain the other. In particular, reconstructing the companion's star-formation history could provide a constraint on the characteristic BH growth timescale.

\item The normalization constrains the seed mass for an adopted critical LW intensity. For $J_{\rm crit}=1000\,J_{21}$, we infer
$\log_{10}(M_{\rm seed}/M_\odot)=6.72^{+0.29}_{-0.39}$, above the range typically predicted for direct-collapse seeds. Lower critical intensities or a systematic downward revision of the BH masses reduce this tension. A uniform mass revision of $-1.5$~dex reduces the inferred seed mass to $\log_{10}(M_{\rm seed}/M_\odot)\simeq5.2$.

\end{itemize}

These results favor a synchronized-pair scenario for the seeding of LRDs with detected companions. Simulations following seed formation and subsequent BH growth can test whether this scenario reproduces the observed slope and scatter, accounting for projection and companion detectability. Dedicated spectroscopic follow-ups will be key to assessing the physical association between the LRD and the UV bright companion, and to robustly measure the SFH of the companion galaxy. At the same time, larger samples with a well-defined selection will test whether the relation persists and tighten constraints on its slope and scatter.

\begin{acknowledgements}
F.Z., S.C, E.P., B.T., S.Z. acknowledge support by European Union’s HE ERC Starting Grant No. 101040227 - WINGS. Views and opinions expressed are however those of the authors only and do not necessarily reflect those of the European Union or the European Research Council Executive Agency. Neither the European Union nor the granting authority can be held responsible for them.
\end{acknowledgements}
\bibliographystyle{aa_url}
\bibliography{biblio}

\begin{thebibliography}{45}
\expandafter\ifx\csname natexlab\endcsname\relax\def\natexlab#1{#1}\fi

\bibitem[{Baccus \& Xu(2026)}]{Baccus2026}
Baccus, C.~J. \& Xu, X. 2026, \href{http://dx.doi.org/10.3847/1538-4357/ae7de7}{\color{magenta}The Astrophysical Journal}, 1006, 165

\bibitem[{Baggen {et~al.}(2026)Baggen, Scoggins, van Dokkum, Haiman, Torralba, \& Matthee}]{Baggen2026}
Baggen, J. F.~W., Scoggins, M.~T., van Dokkum, P., {et~al.} 2026, \href{http://dx.doi.org/10.3847/2041-8213/ae58a5}{\color{magenta}The Astrophysical Journal Letters}, 1002, L4

\bibitem[{{Begelman} \& {Dexter}(2026)}]{Begelman2026}
{Begelman}, M.~C. \& {Dexter}, J. 2026, \href{http://dx.doi.org/10.3847/1538-4357/ae274a}{\color{magenta}\apj}, \href{https://ui.adsabs.harvard.edu/abs/2026ApJ...996...48B}{996, 48}

\bibitem[{Bromm \& Loeb(2003)}]{BrommLoeb2003}
Bromm, V. \& Loeb, A. 2003, \href{http://dx.doi.org/10.1086/377529}{\color{magenta}The Astrophysical Journal}, 596, 34

\bibitem[{{Carnall} {et~al.}(2018){Carnall}, {McLure}, {Dunlop}, \& {Dav{\'e}}}]{Carnall2018}
{Carnall}, A.~C., {McLure}, R.~J., {Dunlop}, J.~S., \& {Dav{\'e}}, R. 2018, \href{http://dx.doi.org/10.1093/mnras/sty2169}{\color{magenta}\mnras}, \href{https://scixplorer.org/abs/2018MNRAS.480.4379C}{480, 4379}

\bibitem[{{Chen} {et~al.}(2025){Chen}, {Ho}, {Li}, \& {Inayoshi}}]{chen2025}
{Chen}, C.-H., {Ho}, L.~C., {Li}, R., \& {Inayoshi}, K. 2025, \href{http://dx.doi.org/10.3847/2041-8213/adee0a}{\color{magenta}\apjl}, \href{https://scixplorer.org/abs/2025ApJ...989L..12C}{989, L12}

\bibitem[{Chon {et~al.}(2016)Chon, Hirano, Hosokawa, \& Yoshida}]{Chon2016}
Chon, S., Hirano, S., Hosokawa, T., \& Yoshida, N. 2016, \href{http://dx.doi.org/10.3847/0004-637X/832/2/134}{\color{magenta}The Astrophysical Journal}, 832, 134

\bibitem[{{de Graaff} {et~al.}(2026){de Graaff}, {Hviding}, {Naidu}, {Greene}, {Miller}, {Leja}, {Matthee}, {Brammer}, {Katz}, {Bezanson}, {Boogaard}, {Bose}, {Chisholm}, {Cleri}, {Dayal}, {Feldmann}, {Fudamoto}, {Fujimoto}, {Furtak}, {Glazebrook}, {Gottumukkala}, {Heintz}, {Kokorev}, {Labbe}, {Maseda}, {McConachie}, {Nanayakkara}, {Nelson}, {Nowaczyk}, {Oesch}, {Rix}, {Setton}, {Torralba}, {Walter}, {Wang}, {Weibel}, \& {van der Wel}}]{deGraaff2026}
{de Graaff}, A., {Hviding}, R.~E., {Naidu}, R.~P., {et~al.} 2026, \href{http://dx.doi.org/10.1093/mnras/stag1567}{\color{magenta}\mnras}, \href{https://ui.adsabs.harvard.edu/abs/2026MNRAS.551g1567D}{551, stag1567}

\bibitem[{Devecchi \& Volonteri(2009)}]{DevecchiVolonteri2009}
Devecchi, B. \& Volonteri, M. 2009, \href{http://dx.doi.org/10.1088/0004-637X/694/1/302}{\color{magenta}The Astrophysical Journal}, 694, 302

\bibitem[{{Ferrara} {et~al.}(2014){Ferrara}, {Salvadori}, {Yue}, \& {Schleicher}}]{Ferrara2014}
{Ferrara}, A., {Salvadori}, S., {Yue}, B., \& {Schleicher}, D. 2014, \href{http://dx.doi.org/10.1093/mnras/stu1280}{\color{magenta}\mnras}, \href{https://ui.adsabs.harvard.edu/abs/2014MNRAS.443.2410F}{443, 2410}

\bibitem[{{Gentile} {et~al.}(2026){Gentile}, {Giavalisco}, {Daddi}, {Elbaz}, {Billand}, {Franco}, {Magnelli}, {Barro}, {Cheng}, {Cleri}, \& et~al.}]{Gentile2026}
{Gentile}, F., {Giavalisco}, M., {Daddi}, E., {et~al.} 2026, \href{http://dx.doi.org/10.1051/0004-6361/202661289}{\color{magenta}\aap}, \href{https://scixplorer.org/abs/2026A&A...714A.463G}{714, A463}

\bibitem[{{Gottumukkala} {et~al.}(2026){Gottumukkala}, {Brammer}, {Toft}, {Xiao}, {de Graaff}, {Oesch}, {Valentino}, {Whitaker}, {Baker}, {Barrufet}, \& et~al.}]{Gottumukkala2026}
{Gottumukkala}, R., {Brammer}, G., {Toft}, S., {et~al.} 2026, \href{https://scixplorer.org/abs/2026arXiv261015901G}{\href{http://dx.doi.org/10.48550/arXiv.2610.15901}{\color{magenta}arXiv e-prints}, arXiv:2610.15901}

\bibitem[{{Greene} \& {Ho}(2005)}]{Greene2005}
{Greene}, J.~E. \& {Ho}, L.~C. 2005, \href{http://dx.doi.org/10.1086/431897}{\color{magenta}\apj}, \href{https://ui.adsabs.harvard.edu/abs/2005ApJ...630..122G}{630, 122}

\bibitem[{Greene {et~al.}(2024)Greene, Labb{\'{e}}, Goulding, Furtak, Chemerynska, Kokorev, Dayal, Volonteri, Williams, Wang, Setton, Burgasser, Bezanson, Atek, Brammer, Cutler, Feldmann, Fujimoto, Glazebrook, de~Graaff, Khullar, Leja, Marchesini, Maseda, Matthee, Miller, Naidu, Nanayakkara, Oesch, Pan, Papovich, Price, van Dokkum, Weaver, Whitaker, \& Zitrin}]{Greene2024}
Greene, J.~E., Labb{\'{e}}, I., Goulding, A.~D., {et~al.} 2024, \href{http://dx.doi.org/10.3847/1538-4357/ad1e5f}{\color{magenta}The Astrophysical Journal}, 964, 39

\bibitem[{{Greene} {et~al.}(2026){Greene}, {Setton}, {Furtak}, {Naidu}, {Volonteri}, {Dayal}, {Labbe}, {van Dokkum}, {Bezanson}, {Brammer}, \& et~al.}]{Greene2025}
{Greene}, J.~E., {Setton}, D.~J., {Furtak}, L.~J., {et~al.} 2026, \href{http://dx.doi.org/10.3847/1538-4357/ae1836}{\color{magenta}\apj}, \href{https://scixplorer.org/abs/2026ApJ...996..129G}{996, 129}

\bibitem[{{Inayoshi}(2025)}]{Inayoshi2025}
{Inayoshi}, K. 2025, \href{http://dx.doi.org/10.3847/2041-8213/adea66}{\color{magenta}\apjl}, \href{https://ui.adsabs.harvard.edu/abs/2025ApJ...988L..22I}{988, L22}

\bibitem[{{Inayoshi} \& {Maiolino}(2025)}]{InayoshiMaiolino2025}
{Inayoshi}, K. \& {Maiolino}, R. 2025, \href{http://dx.doi.org/10.3847/2041-8213/adaebd}{\color{magenta}\apjl}, \href{https://ui.adsabs.harvard.edu/abs/2025ApJ...980L..27I}{980, L27}

\bibitem[{Inayoshi {et~al.}(2020)Inayoshi, Visbal, \& Haiman}]{Inayoshi2020}
Inayoshi, K., Visbal, E., \& Haiman, Z. 2020, \href{http://dx.doi.org/10.1146/annurev-astro-120419-014455}{\color{magenta}Annual Review of Astronomy and Astrophysics}, 58, 27

\bibitem[{{Ji} {et~al.}(2026){Ji}, {Sun}, {Giavalisco}, {Zhu}, {Rieke}, {Williams}, {Maseda}, {Lyu}, {Rieke}, \& {Tacchella}}]{Ji2026}
{Ji}, Z., {Sun}, Y., {Giavalisco}, M., {et~al.} 2026, \href{http://dx.doi.org/10.3847/2041-8213/ae9fea}{\color{magenta}\apjl}, \href{https://scixplorer.org/abs/2026ApJ..1008L..55J}{1008, L55}

\bibitem[{{Jones} {et~al.}(2025){Jones}, {Kocevski}, {Pacucci}, {Taylor}, {Finkelstein}, {Buchner}, {Trump}, {Somerville}, {Hirschmann}, {Yung}, {Barro}, {Bell}, {Bisigello}, {Calabro}, {Cleri}, {Dekel}, {Dickinson}, {Gandolfi}, {Giavalisco}, {Grogin}, {Inayoshi}, {Kartaltepe}, {Koekemoer}, {Napolitano}, {Onoue}, {Ravindranath}, {Rodighiero}, \& {Wilkins}}]{Jones2026}
{Jones}, B.~L., {Kocevski}, D.~D., {Pacucci}, F., {et~al.} 2025, \href{https://ui.adsabs.harvard.edu/abs/2025arXiv251007376J}{\href{http://dx.doi.org/10.48550/arXiv.2510.07376}{\color{magenta}arXiv e-prints}, arXiv:2510.07376}

\bibitem[{{Juod{\v{z}}balis} {et~al.}(2026){Juod{\v{z}}balis}, {Maiolino}, {Baker}, {Lake}, {Scholtz}, {D'Eugenio}, {Trefoloni}, {Isobe}, {Tacchella}, {Bunker}, \& et~al.}]{Juodzbalis2026}
{Juod{\v{z}}balis}, I., {Maiolino}, R., {Baker}, W.~M., {et~al.} 2026, \href{http://dx.doi.org/10.1093/mnras/stag086}{\color{magenta}\mnras}, \href{https://ui.adsabs.harvard.edu/abs/2026MNRAS.546ag086J}{546, stag086}

\bibitem[{Juod{\v{z}}balis {et~al.}(2024)Juod{\v{z}}balis, Ji, Maiolino, D'Eugenio, Scholtz, Risaliti, Fabian, Mazzolari, Gilli, Prandoni, Arribas, Bunker, Carniani, Charlot, Curtis-Lake, de~Graaff, Hainline, Parlanti, Perna, P{\'{e}}rez-Gonz{\'{a}}lez, Robertson, Tacchella, {\"{U}}bler, Williams, Willott, \& Witstok}]{Juodzbalis2024}
Juod{\v{z}}balis, I., Ji, X., Maiolino, R., {et~al.} 2024, \href{http://dx.doi.org/10.1093/mnras/stae2367}{\color{magenta}Monthly Notices of the Royal Astronomical Society}, 535, 853

\bibitem[{{Kennicutt}(1998)}]{Kennicutt1998}
{Kennicutt}, Jr., R.~C. 1998, \href{http://dx.doi.org/10.1146/annurev.astro.36.1.189}{\color{magenta}\araa}, \href{https://scixplorer.org/abs/1998ARA&A..36..189K}{36, 189}

\bibitem[{Kocevski {et~al.}(2023)Kocevski, Onoue, Inayoshi, Trump, Arrabal~Haro, Grazian, Dickinson, Finkelstein, Kartaltepe, Hirschmann, Aird, Holwerda, Fujimoto, Juneau, Amor{\'i}n, Backhaus, Bagley, Barro, Bell, Bisigello, Calabr{\`o}, Cleri, Cooper, Ding, Grogin, Ho, Hutchison, Inoue, Jiang, Jones, Koekemoer, Li, Li, McGrath, Molina, Papovich, P{\'e}rez-Gonz{\'a}lez, Pirzkal, Wilkins, Yang, \& Yung}]{Kocevski2023}
Kocevski, D.~D., Onoue, M., Inayoshi, K., {et~al.} 2023, \href{http://dx.doi.org/10.3847/2041-8213/ace5a0}{\color{magenta}The Astrophysical Journal Letters}, 954, L4

\bibitem[{{Kokorev} {et~al.}(2026){Kokorev}, {Chisholm}, {Naidu}, {Fujimoto}, {Atek}, {Brammer}, {Finkelstein}, {Akins}, {Berg}, {Furtak}, {Fei}, {Hsiao}, {Labb{\'e}}, {Matthee}, {Mu{\~n}oz}, {Oesch}, {Pan}, {Rinaldi}, {Saldana-Lopez}, {Schaerer}, {Volonteri}, \& {Zitrin}}]{Kokorev2026}
{Kokorev}, V., {Chisholm}, J., {Naidu}, R.~P., {et~al.} 2026, \href{http://dx.doi.org/10.3847/1538-4357/ae4ed7}{\color{magenta}\apj}, \href{https://ui.adsabs.harvard.edu/abs/2026ApJ..1004..153K}{1004, 153}

\bibitem[{{Leja} {et~al.}(2019){Leja}, {Carnall}, {Johnson}, {Conroy}, \& {Speagle}}]{Leja2019}
{Leja}, J., {Carnall}, A.~C., {Johnson}, B.~D., {Conroy}, C., \& {Speagle}, J.~S. 2019, \href{http://dx.doi.org/10.3847/1538-4357/ab133c}{\color{magenta}\apj}, \href{https://scixplorer.org/abs/2019ApJ...876....3L}{876, 3}

\bibitem[{Madau {et~al.}(2014)Madau, Haardt, \& Dotti}]{Madau2014}
Madau, P., Haardt, F., \& Dotti, M. 2014, \href{http://dx.doi.org/10.1088/2041-8205/784/2/L38}{\color{magenta}The Astrophysical Journal Letters}, 784, L38

\bibitem[{Madau \& Rees(2001)}]{MadauRees2001}
Madau, P. \& Rees, M.~J. 2001, \href{http://dx.doi.org/10.1086/319848}{\color{magenta}The Astrophysical Journal}, 551, L27

\bibitem[{{Maiolino} {et~al.}(2024){Maiolino}, {Scholtz}, {Curtis-Lake}, {Carniani}, {Baker}, {de Graaff}, {Tacchella}, {{\"U}bler}, {D'Eugenio}, {Witstok}, \& et~al.}]{Maiolino2024}
{Maiolino}, R., {Scholtz}, J., {Curtis-Lake}, E., {et~al.} 2024, \href{http://dx.doi.org/10.1051/0004-6361/202347640}{\color{magenta}\aap}, \href{https://ui.adsabs.harvard.edu/abs/2024A&A...691A.145M}{691, A145}

\bibitem[{Matthee {et~al.}(2024)Matthee, Naidu, Brammer, Chisholm, Eilers, Goulding, Greene, Kashino, Labb{\'{e}}, Lilly, Mackenzie, Oesch, Weibel, Wuyts, Xiao, Bordoloi, Bouwens, van Dokkum, Illingworth, Kramarenko, Maseda, Mason, Meyer, Nelson, Reddy, Shivaei, Simcoe, \& Yue}]{Matthee2024}
Matthee, J., Naidu, R.~P., Brammer, G., {et~al.} 2024, \href{http://dx.doi.org/10.3847/1538-4357/ad2345}{\color{magenta}The Astrophysical Journal}, 963, 129

\bibitem[{{Modelevsky} {et~al.}(2026){Modelevsky}, {Nitzan}, {Sari}, \& {Quataert}}]{Modelevsky2026}
{Modelevsky}, E., {Nitzan}, O., {Sari}, R., \& {Quataert}, E. 2026, \href{https://ui.adsabs.harvard.edu/abs/2026arXiv260805283M}{\href{http://dx.doi.org/10.48550/arXiv.2608.05283}{\color{magenta}arXiv e-prints}, arXiv:2608.05283}

\bibitem[{{Naidu} {et~al.}(2026){Naidu}, {Matthee}, {Katz}, {de Graaff}, {Oesch}, {Smith}, {Greene}, {Brammer}, {Weibel}, {Hviding}, {Chisholm}, {Labb{\'e}}, {Simcoe}, {Witten}, {Sun}, {Atek}, {Baggen}, {Belli}, {Bezanson}, {Boogaard}, {Bose}, {Bouwens}, {Covelo-Paz}, {Dayal}, {Fudamoto}, {Furtak}, {Giovinazzo}, {Goulding}, {Gronke}, {Heintz}, {Hirschmann}, {Illingworth}, {Inoue}, {Johnson}, {Leja}, {Leonova}, {McConachie}, {Maseda}, {Natarajan}, {Nelson}, {Setton}, {Shivaei}, {Sobral}, {Stefanon}, {Tacchella}, {Toft}, {Torralba}, {van Dokkum}, {van der Wel}, {Volonteri}, {Walter}, {Wang}, {Watson}, \& {Whitaker}}]{Naidu2026}
{Naidu}, R.~P., {Matthee}, J., {Katz}, H., {et~al.} 2026, \href{http://dx.doi.org/10.1038/s41586-026-10846-4}{\color{magenta}\nat}, \href{https://ui.adsabs.harvard.edu/abs/2026Natur.656..329N}{656, 329}

\bibitem[{{Pacucci} {et~al.}(2026){Pacucci}, {Ferrara}, \& {Kocevski}}]{Pacucci_Ferrara2026}
{Pacucci}, F., {Ferrara}, A., \& {Kocevski}, D.~D. 2026, \href{https://ui.adsabs.harvard.edu/abs/2026arXiv260114368P}{\href{http://dx.doi.org/10.48550/arXiv.2601.14368}{\color{magenta}arXiv e-prints}, arXiv:2601.14368}

\bibitem[{Pacucci \& Rinaldi(2026)}]{PacucciRinaldi2026}
Pacucci, F. \& Rinaldi, P. 2026, \href{http://dx.doi.org/10.48550/arXiv.2610.18968}{\color{magenta}arXiv e-prints}, arXiv:2610.18968

\bibitem[{Regan {et~al.}(2017)Regan, Visbal, Wise, Haiman, Johansson, \& Bryan}]{Regan2017}
Regan, J.~A., Visbal, E., Wise, J.~H., {et~al.} 2017, \href{http://dx.doi.org/10.1038/s41550-017-0075}{\color{magenta}Nature Astronomy}, 1, 0075

\bibitem[{Rusakov {et~al.}(2026)Rusakov, Watson, Nikopoulos, Brammer, Gottumukkala, Harvey, Heintz, Damgaard, Sim, Sneppen, Vijayan, Adams, Austin, Conselice, Goolsby, Toft, \& Witstok}]{Rusakov2026}
Rusakov, V., Watson, D., Nikopoulos, G.~P., {et~al.} 2026, \href{http://dx.doi.org/10.1038/s41586-025-09900-4}{\color{magenta}Nature}, 649, 574

\bibitem[{{Salim} {et~al.}(2018){Salim}, {Boquien}, \& {Lee}}]{Salim2018}
{Salim}, S., {Boquien}, M., \& {Lee}, J.~C. 2018, \href{http://dx.doi.org/10.3847/1538-4357/aabf3c}{\color{magenta}\apj}, \href{https://scixplorer.org/abs/2018ApJ...859...11S}{859, 11}

\bibitem[{{Schaerer} {et~al.}(2025){Schaerer}, {Guibert}, {Marques-Chaves}, \& {Martins}}]{Schaerer2025}
{Schaerer}, D., {Guibert}, J., {Marques-Chaves}, R., \& {Martins}, F. 2025, \href{http://dx.doi.org/10.1051/0004-6361/202451454}{\color{magenta}\aap}, \href{https://scixplorer.org/abs/2025A&A...693A.271S}{693, A271}

\bibitem[{{Scoggins} \& {Haiman}(2024)}]{ScogginsHaiman2024}
{Scoggins}, M.~T. \& {Haiman}, Z. 2024, \href{http://dx.doi.org/10.1093/mnras/stae1449}{\color{magenta}\mnras}, \href{https://ui.adsabs.harvard.edu/abs/2024MNRAS.531.4584S}{531, 4584}

\bibitem[{Shang {et~al.}(2010)Shang, Bryan, \& Haiman}]{Shang2010}
Shang, C., Bryan, G.~L., \& Haiman, Z. 2010, \href{http://dx.doi.org/10.1111/j.1365-2966.2009.15960.x}{\color{magenta}Monthly Notices of the Royal Astronomical Society}, 402, 1249

\bibitem[{Taylor {et~al.}(2025)Taylor, Finkelstein, Kocevski, Jeon, Bromm, Amor{\'{i}}n, Arrabal~Haro, Backhaus, Bagley, Ba{\~{n}}ados, Bhatawdekar, Brooks, Calabr{\`{o}}, Ch{\'{a}}vez~Ortiz, Cheng, Cleri, Cole, Davis, Dickinson, Donnan, Dunlop, Ellis, Fern{\'{a}}ndez, Fontana, Fujimoto, Giavalisco, Grazian, Guo, Hathi, Holwerda, Hirschmann, Inayoshi, Kartaltepe, Khusanova, Koekemoer, Kokorev, Larson, Leung, Lucas, McLeod, Napolitano, Onoue, Pacucci, Papovich, P{\'{e}}rez-Gonz{\'{a}}lez, Pirzkal, Somerville, Trump, Wilkins, Yung, \& Zhang}]{Taylor2025}
Taylor, A.~J., Finkelstein, S.~L., Kocevski, D.~D., {et~al.} 2025, \href{http://dx.doi.org/10.3847/1538-4357/add15b}{\color{magenta}The Astrophysical Journal}, 986, 165

\bibitem[{Visbal {et~al.}(2014)Visbal, Haiman, \& Bryan}]{Visbal2014}
Visbal, E., Haiman, Z., \& Bryan, G.~L. 2014, \href{http://dx.doi.org/10.1093/mnras/stu1794}{\color{magenta}Monthly Notices of the Royal Astronomical Society}, 445, 1056

\bibitem[{{Whalen} {et~al.}(2026){Whalen}, {Latif}, {Topalakis}, {Cullen}, {Nandal}, \& {Khochfar}}]{Whalen2026}
{Whalen}, D.~J., {Latif}, M.~A., {Topalakis}, K., {et~al.} 2026, \href{http://dx.doi.org/10.3847/2041-8213/aeb37a}{\color{magenta}\apjl}, \href{https://ui.adsabs.harvard.edu/abs/2026ApJ..1011L..29W}{1011, L29}

\bibitem[{{Yue} {et~al.}(2017){Yue}, {Ferrara}, {Pacucci}, \& {Omukai}}]{Yue2017}
{Yue}, B., {Ferrara}, A., {Pacucci}, F., \& {Omukai}, K. 2017, \href{http://dx.doi.org/10.3847/1538-4357/aa6627}{\color{magenta}\apj}, \href{https://ui.adsabs.harvard.edu/abs/2017ApJ...838..111Y}{838, 111}

\bibitem[{{Ziparo} {et~al.}(2026){Ziparo}, {Carniani}, {Gallerani}, \& {Trefoloni}}]{Ziparo2026}
{Ziparo}, F., {Carniani}, S., {Gallerani}, S., \& {Trefoloni}, B. 2026, \href{http://dx.doi.org/10.1051/0004-6361/202659574}{\color{magenta}\aap}, \href{https://scixplorer.org/abs/2026A&A...714A.290Z}{714, A290}

\end{thebibliography}

\appendix

\section{Sample and observables}
\label{app:sample}
We analyze 24 LRD systems drawn from the companion catalogue of \citet{Baggen2026} plus one from Zamora et al. \textit{in prep.}, with BH mass estimates compiled from \citet{Taylor2025}, \citet{Greene2024}, \citet{Matthee2024}, and \citet{Baccus2026}. The adopted properties, alternative source identifiers and mass references are listed in Table~\ref{tab:sample}.

We adopt the effective projected separations, $d_{\rm eff}$, and total companion UV absolute magnitudes, $M_{\rm UV,tot}$, reported by \citet{Baggen2026}. For systems with multiple companions, $d_{\rm eff}$ is defined through $L_{\nu,\rm LW}^{\rm tot}/d_{\rm eff}^{2}
=\sum_i L_{\nu,\rm LW,i}/d_i^{2}$, where $d_i$ and $L_{\nu,\rm LW,i}$ are the projected separation and LW specific luminosity of each companion. It therefore represents the separation at which a single source with the combined LW luminosity would reproduce the summed
radiation field of all companions.

We convert $M_{\rm UV,tot}$ into the total rest-frame specific luminosity at $1500\,\text{\AA}$, $L_{\rm UV,obs}$, using the AB magnitude definition. Assuming a common LW-to-UV specific luminosity ratio for all companions, the same effective separation also reproduces their combined UV radiation field. We therefore estimate the incident UV specific intensity at the LRD location as
$J_{\rm UV}=L_{\rm UV,obs}/(16\pi^2d_{\rm eff}^{2})$. For lensed systems, we use the magnification-corrected companion luminosities and separations reported by \citet{Baggen2026}.

\begin{table*}[!b]
\caption{Properties of the systems used in the correlation analysis.}
\label{tab:sample}
\centering
\small
\setlength{\tabcolsep}{4pt}
\begin{tabular}{lrrrrrll}
\hline\hline
Source & $z$ & $d_{\rm eff}$ & $M_{\rm UV,tot}$ &
$\log_{10}(M_{\rm BH}/M_\odot)$ & $\sigma_{\rm lit}$ &
Mass-catalogue ID & Ref. \\
 & & [kpc] & [AB mag] & & [dex] & & \\
\hline
CEERS-19799   & 4.22 & 2.42 & -20.2 & 7.35 &
$0.04$ & RUBIES-EGS-28812 & T \\
CEERS-71365   & 4.80 & 1.75 & -19.4 & 5.79 &
${}^{+0.027}_{-0.030}$ & EGS\_4106\_71365 & B \\
CEERS-7902    & 6.99 & 0.87 & -20.2 & 7.85 &
$0.01$ & RUBIES\_4233\_55604 & B \\
CEERS-10444   & 6.69 & 0.52 & -19.1 & 8.26 &
${}^{+0.02}_{-0.01}$ & RUBIES-EGS-49140 & T \\
CEERS-13318   & 5.28 & 1.89 & -19.9 & 8.30 &
$0.02$ & RUBIES-EGS-42046 & T \\
UNCOVER-4286  & 5.84 & 0.37 & -18.5 & 8.00 &
$0.30$ & 4286 & G \\
UNCOVER-13821 & 6.34 & 0.44 & -16.2 & 7.30 &
$\ldots$ & 13821 & G \\
UNCOVER-23608 & 5.80 & 0.79 & -18.6 & 7.50 &
$0.20$ & 23608 & G \\
UNCOVER-35488 & 6.26 & 0.31 & -18.1 & 7.40 &
$0.20$ & 35488 & G \\
UNCOVER-38108 & 4.96 & 2.80 & -18.4 & 8.40 &
$0.50$ & 38108 & G \\
UNCOVER-41225 & 6.76 & 0.68 & -18.2 & 7.70 &
$0.40$ & 41225 & G \\
GOODS-S-13971 & 5.48 & 2.21 & -19.7 & 6.88 &
$0.01$ & JADES\_1286\_204851 & B \\
JADES-13329   & 3.94 & 3.95 & -19.0 & 6.30 &
${}^{+0.038}_{-0.039}$ & JADES\_1180\_13329 & B \\
JADES-68797   & 5.04 & 0.72 & -20.4 & 7.82 &
$0.01$ & JADES\_1181\_68797 & B \\
JADES-73488   & 4.13 & 1.07 & -17.2 & 6.84 &
$0.01$ & JADES\_1181\_73488 & B \\
J1148-7111    & 4.34 & 0.92 & -19.6 & 7.92 &
$0.10$ & J1148-7111 & M \\
J1148-21787   & 4.28 & 1.23 & -19.3 & 7.59 &
$0.18$ & J1148-21787 & M \\
J0100-2017    & 4.94 & 1.49 & -17.6 & 7.44 &
$0.11$ & J0100-2017 & M \\
J0100-12446   & 4.70 & 1.49 & -19.5 & 7.46 &
$0.06$ & J0100-12446 & M \\
J0100-15157   & 4.94 & 0.86 & -20.6 & 7.35 &
$0.08$ & J0100-15157 & M \\
J0100-16221   & 4.35 & 1.40 & -19.1 & 7.53 &
$0.07$ & J0100-16221 & M \\
J0148-4214    & 5.02 & 0.53 & -20.1 & 7.32 &
$0.10$ & J0148-4214 & M \\
J0148-12884   & 4.60 & 1.77 & -20.1 & 6.91 &
$0.15$ & J0148-12884 & M \\
JADES-28074   & 2.27 & 0.30 & -19.6 & 8.26 &
$0.30$ & JADES\_1181\_28074 & J \\
\hline
\end{tabular}

\tablefoot{
$\sigma_{\rm lit}$ denotes the uncertainty quoted for the matching
literature mass estimate. The fit adopts uniform uncertainties of
$0.4$~dex in $\log_{10}(M_{\rm BH}/M_\odot)$ and $0.3$~dex in
$\log_{10}J_{\rm UV}$.
Companion magnitudes refer to the combined UV emission of all
identified companions.
Redshifts, magnitudes and separations follow \citet{Baggen2026}.
References:
T = \citet{Taylor2025}, Table~1;
G = \citet{Greene2024}, Table~3;
M = \citet{Matthee2024}, Table~3;
B = \citet{Baccus2026}, Table~1;
J = \citet{Juodzbalis2024}, Sect.~5.1.
}
\end{table*}

\section{An exponential description of the star-formation histories}
\label{app:sfh_tau}
Here we justify the choice of an exponential star-formation history (SFH) as an approximation of the SFHs in our sample.
To this end, we use the SFHs from the DAWN \textit{JWST} Archive (DJA) catalogue of \citet{Gottumukkala2026} (\texttt{djapipes} v4.6.0). They fitted the \textit{JWST}/NIRSpec PRISM spectra together with the NIRCam broad-band photometry, using a modified version of \textsc{bagpipes} \citep{Carnall2018} with the \textsc{nautilus} nested sampler, with a narrow redshift prior centred on the spectroscopic value. The stellar populations are BPASS models, with nebular emission computed with \textsc{cloudy}. The dust attenuation follows a flexible \citet{Salim2018} law, with extra attenuation for stars younger than 25\,Myr. A polynomial calibration and a noise-scaling term account for differences between the spectrum and the photometry. The SFH is non-parametric, with the ``continuity'' prior \citep{Leja2019}: the SFR is constant within fixed bins of lookback time, and the change in $\log{\rm SFR}$ between adjacent bins has a Student's-$t$ prior (scale 0.3\,dex, $\nu=2$), which favours smooth SFHs but allows sharp changes. The bin edges are at 0, 3, 10, 25, 50 and 100\,Myr, followed by logarithmically spaced bins. For each bin we take the posterior median SFR and the 16th--84th percentile range.

From the 4401 galaxies with $4<z<7$, we select the 3119 with $M_{\rm UV}>-19.5$.
For each galaxy we fit
\begin{equation}
  {\rm SFR}(t) = {\rm SFR}_0\, \exp(-t/\tau_\star),
  \label{eq:exp_sfh}
\end{equation}
where $t$ is the lookback time from the epoch of observation, so ${\rm SFR}_0$ is the SFR at
observation.
The fit minimizes $\chi^2$ in $\log{\rm SFR}$. All but 72 galaxies (2.3\%) have $\tau_\star>0$. We exclude these few declining SFHs and keep 3047 galaxies.

At these redshifts the SFHs rise steeply and exponential profiles provide a reasonable description of the reconstructed SFHs. In particular, the fit is statistically acceptable ($p>0.05$,
where $p$ is the $p$-value of the $\chi^2$ goodness-of-fit test)
for 70\% of the galaxies, with a median reduced $\chi^2$ of 1.3.

The distribution of $\tau_\star$ is close to log-normal,
with median $\tau_\star = 98$\,Myr, 16th--84th percentile range 50--251\,Myr and
$\sigma(\log\tau_\star)=0.38$\,dex.

\section{Testing external photoionization of the UV companions}
\label{app:photoionization}

The spatially offset UV emission could arise from gas illuminated by the
LRD. If the UV continuum is dominated by nebular emission powered by the
LRD, we expect its luminosity to increase with the incident ionizing flux
for clouds with comparable intercepting areas and reprocessing
efficiencies. We test this expectation in our sample of 25 systems, using the 23
with available LRD H$\alpha$ luminosities.

We use the H$\alpha$ luminosity of the LRD and the effective projected
separation $d_{\rm eff}$ to construct a proxy for the radiation field
incident on the companion,
\begin{equation}
F_{\rm H\alpha,LRD} =
\frac{L_{\rm H\alpha,LRD}}{4\pi d_{\rm eff}^{\,2}}.
\end{equation}
We compare this proxy with the total rest-frame UV luminosity of the
companions, $\lambda L_{\lambda,\rm UV}=\nu L_{\nu,\rm UV}$ at
$1500\,\text{\AA}$, derived from their absolute UV magnitudes
\citep{Baggen2026}.

\begin{figure}
    \centering
    \includegraphics[width=\columnwidth]{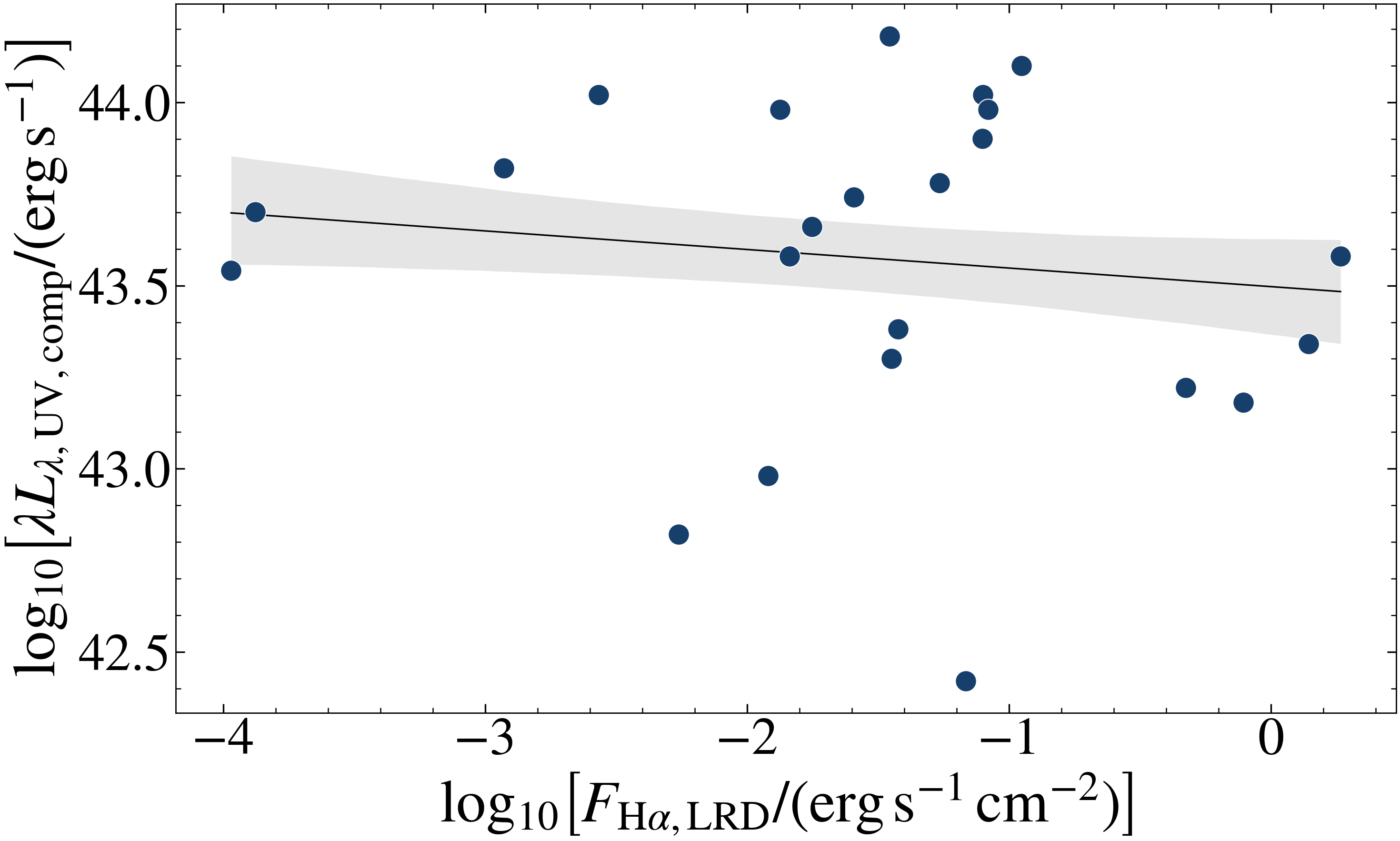}
    \caption{Total companion UV luminosity at $1500\,\text{\AA}$ as a
    function of the incident H$\alpha$ flux proxy from the LRD. The black line shows the
    unweighted fit in log space and the grey band marks the
    16th--84th percentiles of the bootstrap fits at each flux.}
    \label{fig:halpha_companion_uv}
\end{figure}

Figure~\ref{fig:halpha_companion_uv} shows that the companion UV
luminosity is consistent with a flat relation over more than four
orders of magnitude in $F_{\rm H\alpha,LRD}$. An unweighted linear fit
in log space gives a slope of $-0.051$, with a 95\% interval of
$[-0.177,0.071]$ from $10^4$ bootstrap resamples of the systems.
The Pearson and Spearman coefficients are $r=-0.124$ and
$\rho_{\rm S}=-0.052$, with two-sided permutation probabilities
of $0.570$ and $0.811$, respectively, based on $2\times10^5$
permutations.

The absence of a rising trend disfavors a simple model in which the
UV luminosities of the companions are set primarily by reprocessing of
radiation from the LRD under comparable gas conditions. We use the LRD H$\alpha$ luminosity as an approximate proxy for its bolometric luminosity \citep{Greene2025, deGraaff2026}. This interpretation depends on how well the LRD H$\alpha$ luminosity traces the ionizing radiation that reaches each companion. Variations in ionizing escape fraction, cloud geometry, and the difference between projected and physical separations could weaken the expected trend.

Finally we also compare the ionizing photon budget inferred from the LRD H$\alpha$ luminosity with that required to power the nebular emission in the companion galaxy. We assume ionization equilibrium, case-B recombination, and negligible dust absorption. Adopting the standard H$\alpha$ conversion at $T_e = 10^4$~K \citep{Kennicutt1998}, the ionizing photon rate absorbed by the LRD line-emitting gas is $Q_{\rm abs,LRD}=7.3\times10^{53}\,{\rm s^{-1}}\left[L_{\rm H\alpha}/(10^{42}\,{\rm erg\,s^{-1}})\right]$.

This measures the absorbed photon rate rather than the total production rate. Defining $f_{\rm esc}$ as the fraction of ionizing photons escaping the LRD, the total production rate is
$Q_{\rm src} = Q_{\rm abs,LRD}/(1-f_{\rm esc})$.

For the companion, we assume that its entire rest-frame 1500~\AA{} continuum is nebular. \citet{Schaerer2025} report
pure-nebular ionizing efficiencies of $\log_{10}[\xi_{\rm ion}^{\rm neb}/({\rm Hz\,erg^{-1}})]\simeq 26.34$ for electron temperatures of $10^4$~K. This conversion includes the nebular free-bound, free-free, and two-photon continua. The required absorbed photon rate is therefore $Q_{\rm req} = L_{\rm UV}\xi_{\rm ion}^{\rm neb}$. 

Assuming isotropic escaping radiation, a necessary
condition for externally powered nebular emission is consequently $Q_{\rm req}\leq f_\Omega\,\frac{f_{\rm esc}}{1-f_{\rm esc}}Q_{\rm abs,LRD}$  where  $f_\Omega$  is the fraction of  solid angle subtended by the companion.
With these conversions and an illustrative covering fraction $f_\Omega = 0.25$, we find minimum escape fractions ranging from approximately 18\% to 96\%, with 9 systems requiring $f_{\rm esc}>90\%$. Under the adopted assumption of isotropic escape, such high fractions would be difficult to reconcile with models in which the central ionizing source is surrounded by a dense, optically thick gas envelope that absorbs most of its ionizing radiation.

\section{Statistical checks of the mass--intensity relation}
\label{app:robustness}

We examine the mass--intensity relation using complementary statistical procedures. Throughout this appendix,
$x_i=\log_{10}[J_{{\rm UV},i}/ ({\rm erg\,s^{-1}\,cm^{-2}\,Hz^{-1}\,sr^{-1}})]$ and $y_i=\log_{10}(M_{{\rm BH},i}/M_\odot)$. All 25 systems enter the analysis.

We fit the relation $y=\beta+\alpha x$ using an orthogonal Bayesian regression, adopting Gaussian uncertainties of $0.3$~dex in $x$ and $0.4$~dex in $y$. The fit uses the distances perpendicular to the line, with the uncertainties in both coordinates projected onto
this direction. An intrinsic perpendicular scatter, $\sigma_\perp$, accounts for dispersion beyond the adopted measurement uncertainties.
We adopt uniform priors on the line angle and perpendicular offset, and a half-normal prior on $\sigma_\perp$ with a scale of $1$~dex. The posterior medians and 16th--84th percentile intervals are $\alpha=0.88^{+0.40}_{-0.28}$ and $\beta=22.34^{+6.80}_{-4.83}$.
The fit does not explicitly model the three-dimensional separations or the sample selection.

We also evaluate the association using correlation coefficients. The measured data have Pearson coefficient $r=0.400$ and Spearman coefficient $\rho_{\rm S}=0.381$. We assess the significance of the Spearman coefficient by randomly permuting the BH mass ranks relative to the fixed UV intensity ranks $B=10^6$ times. This preserves the measured values while randomizing their association. We count permutations with an absolute correlation coefficient at least as large as the observed value and estimate the two-sided permutation probability as
\begin{equation}
p_{\rm perm,S}=
\frac{1+\#\{|\rho_{{\rm S},j}|\geq|\rho_{\rm S,obs}|\}}
{B+1}.
\end{equation}
There are 61158 exceedances, giving $p_{\rm perm,S}=6.116\times10^{-2}$. The 95\% binomial interval for the simulated tail probability
is $[6.069,6.163]\times10^{-2}$. This interval describes Monte Carlo precision. The permutation test uses the measured ranks and does not
propagate measurement errors.

As a separate significance test, we compare a free-slope relation with a flat relation ($\alpha=0$). For this comparison, we hold the measured UV intensities fixed and fit Gaussian residuals in BH mass, including the adopted $0.4$~dex mass uncertainties and an intrinsic-scatter term. The intercept and intrinsic scatter are fitted independently in each model, and the free slope can have either sign.

We quantify the improvement over the flat relation using
\begin{equation}
T=2\left[
\ln\mathcal{L}_{\rm free,max}
-\ln\mathcal{L}_{\rm flat,max}
\right].
\end{equation}
To calibrate this statistic, we generate mock BH masses from the best-fitting flat model, retaining the observed UV intensities and mass uncertainties. We refit both models to each mock sample and calculate the fraction yielding $T\geq T_{\rm obs}$. This parametric bootstrap gives $p_{\rm boot}=0.048$. Unlike the primary orthogonal regression, this test is conditional on the measured UV intensities and does not propagate their uncertainties. To assess the robustness of the relation to the possible spectroscopic rejection of some candidate associations, we repeat the orthogonal Bayesian fit for $10^4$ random subsamples, each obtained by removing 7 of the 25 systems
($28\%$). This tests whether the trend persists after excluding approximately $30\%$ of the sample. The posterior median slope is positive in 9,989 trials. Across subsamples, the median of these slopes is $0.92$, with 16th--84th percentiles of $[0.59,1.14]$. These percentiles describe variation among subsamples rather than the posterior uncertainty for the full sample.

\end{document}